\pdfoutput=1

\documentclass[a4paper,11pt]{article}
\usepackage{jcappub} 
\usepackage{color}
\usepackage{xcolor}
\usepackage{amsmath}
\usepackage{amsfonts}
\usepackage{graphicx} 
\usepackage{amssymb}
\usepackage{epstopdf}
\usepackage{url}
\usepackage{bm}
\usepackage{hhline}
\usepackage{subfigure}
\usepackage{multirow}
\usepackage{epsfig}
\usepackage{varioref}
\usepackage{amssymb}
\usepackage{comment} 
\usepackage{cases}
\usepackage{multirow}
\usepackage{mathrsfs}
\usepackage{calrsfs}
\usepackage{mathtools}
\usepackage{amsthm}
\usepackage{hyperref}

\include{configs} 

\title{The observational position of simple non-minimally coupled inflationary scenarios}
\author{David C. Edwards and Andrew R. Liddle} 
\affiliation{Institute for Astronomy, University of Edinburgh, Royal Observatory, Edinburgh EH9~3HJ, United Kingdom}

\emailAdd{dce@roe.ac.uk}
\emailAdd{arl@roe.ac.uk}
\date{\today}

\abstract{We consider two classes of non-minimally coupled inflation models; those with a quadratic coupling of the inflaton to gravity, and the `Universal Attractor' models where the coupling is connected to the potential. We make a detailed analysis of the attractor structure in the latter case, identifying a shift of the attractor from the Starobinsky point and determining conditions for approach to the attractor. We then assess the viability of the models under different interpretations of the BICEP2 experiment's detection of B-mode polarisation, finding strong constraints on the non-minimal coupling in the case where the B-modes have mostly primordial origin.}

\begin{document}
\maketitle

\section{Introduction}

The strong upper limit set on the tensor-to-scalar ratio $r$ by the {\it Planck} satellite mission \cite{PlanckInflation} prompted considerable interest in inflationary models predicting low values of $r$. Amongst those are models where $r$ is of order $(1-n_{\rm s})^2$, where $n_{\rm s}$ is the scalar spectral index, rather than the more common linear relation. This class includes the original $R^2$, or Starobinsky, model of inflation \cite{Starobinsky80}, the non-minimally coupled Higgs inflation model \cite{Bezrukov08}, and a set of models motivated by superconformal symmetry often referred to as `Universal Attractor' models \cite{LKOct13}. These models turn out to be closely related to each other \cite{Kehagias13}. The first part of this article makes a detailed investigation of the attractor structure of these models through both analytic and numerical calculation.

The observational situation has recently sharpened considerably with the detection of B-mode polarisation by the BICEP2 experiment \cite{BICEP2}, which presently has competing interpretations as a consequence of polarised emission from dust \cite{BICEP2,Mortonson14,Flauger14} or as due to primordial tensor perturbations. While the former interpretation leaves the situation unchanged from {\it Planck} \cite{Mortonson14}, the latter acts strongly against the models we are considering and would for the first time impose a meaningful constraint on the magnitude of the non-minimal coupling. The second part of this article explores the constraints under different interpretations of the BICEP2 data, considering also models where the non-minimal coupling is of quadratic form.

\section{The Universal Attractor models}

\subsection{The models}

Recently a Lagrangian
\begin{equation}
	\label{eq:jordanFrameLKOct13}
	\mathcal{L}_{\rm J} = \sqrt{-g}\left[\frac{1}{2}\Omega(\phi)R-\frac{1}{2}(\partial\phi)^2-V_{\rm J}(\phi)\right] \,,
\end{equation}
with
\begin{equation}
	\Omega(\phi) = 1+\xi f(\phi) \quad ; \quad V_{\rm J}(\phi) = \lambda^2f^2(\phi) \,,
\end{equation}
featuring a non-minimal coupling between the inflaton field, $\phi$, and the scalar curvature, $R$, has been suggested as a viable candidate for inflation \cite{LKOct13}. In this section $\xi$ is always taken to be positive (in this sign convention), which is necessary to get Universal Attractor behaviour, and here and throughout the reduced Planck mass is set to unity.
The motivation for such models is rooted in superconformal theories and, while there are `naturalness' problems in the connection between the coupling term and Jordan frame potential \cite{Kehagias13}, under a certain set of conditions they all imply the same scalar spectral index and tensor-to-scalar ratio, regardless of the function $f(\phi)$. Moreover,  the combined values of those are well placed within the {\it Planck} contours in the $n_{\rm s}$--$r$ plane \cite{PlanckInflation}.

The Jordan frame Lagrangian, equation (\ref{eq:jordanFrameLKOct13}), can be transformed to the Einstein frame by use of the conformal transformation:
\begin{equation}
	\label{eq:conformalTransform}
	g_{\mu\nu} \rightarrow \Omega^{-1}(\phi)g_{\mu\nu} \,,
\end{equation}
which gives the Einstein frame Lagrangian
\begin{equation}
	\label{eq:einsteinFrameLKOct13}
	\mathcal{L}_{\rm E} = \sqrt{-g}\left[\frac{1}{2}R-\frac{1}{2\Omega^2(\phi)}\left(\Omega(\phi)+\frac{3}{2}\left[\frac{d\Omega(\phi)}{d\phi}\right]^2\right)(\partial\phi)^2-\frac{V_{\rm J}(\phi)}{\Omega^2(\phi)}\right]\,.
\end{equation}
It is then desirable to make a field transformation to obtain a Lagrangian with a canonical kinetic term  in the Einstein frame, so that all the inflationary dynamics are contained in the potential as in the single scalar field case. This required transformation is
\begin{equation}
	\label{eq:canonicalScalarFieldTransform}
	\left[\frac{\partial\varphi}{\partial\phi}\right]^2 = \frac{1}{\Omega^2(\phi)}\left(\Omega(\phi)+\frac{3}{2}\left[\frac{d\Omega(\phi)}{d\phi}\right]^2\right) \,.
\end{equation}
This allows the potential slow-roll parameters \cite{Liddle:1992wi} to be calculated as
\begin{equation}
	\label{eq:slowRollParametersLKOct13}
	\epsilon \equiv \frac{\Omega^4(\phi)}{V_{\rm J}^2(\phi)}\left(\frac{d}{d\phi}\left[\frac{V_{\rm J}(\phi)}{\Omega^2(\phi)}\right]\right)^2\left(\frac{\partial\phi}{\partial\varphi}\right)^2 \quad ; \quad   \eta \equiv \frac{\Omega^2(\phi)}{V_{\rm J}(\phi)}\frac{\partial\phi}{\partial\varphi}\frac{d}{d\phi}\left(\frac{\partial\phi}{\partial\varphi}\frac{d}{d\phi}\left[\frac{V_{\rm J}(\phi)}{\Omega^2(\phi)}\right]\right) \,.
\end{equation}

By utilizing these expressions in the usual expression for the number of $e$-foldings it is possible to show \citep{LKOct13} attractor behaviour at strong coupling, where $\xi f(\phi) \gg 1 $. The first slow-roll parameter is given in terms of the function $f(\phi)$ by
\begin{equation}
	\label{eq:epsilonLKOct2013}
	\epsilon = \left[\frac{2f'(\phi)}{f(\phi)}\right]^2\frac{1}{(2+2\xi f(\phi)+ 3[\xi f'(\phi)]^2)} \,.
\end{equation}
Then the number of $e$-foldings is
\begin{equation}
	\label{eq:fullN}
	N = \int \frac{1}{\sqrt{2\epsilon}}d\varphi = \frac{1}{4}\int \frac{f(\phi)}{f'(\phi)}\frac{2+2\xi f(\phi)+3[\xi f'(\phi)]^2}{1+\xi f(\phi)}d\phi \,.
\end{equation}
In the high-coupling limit $N$ is given by
\begin{equation}
	\label{eq:numberOfEFoldingsStrongCouplingLKOct13}
	N = \int^{\phi_N}_{\phi_{\text{end}}}\left(\frac{3}{4}\xi f'(\phi)+\frac{f(\phi)}{2f'(\phi)}-\frac{3f'(\phi)}{4f(\phi)}\right)d\phi \,,
\end{equation}
which can then be simplified further. The explicit conditions for the simplification are most easily seen when equation (\ref{eq:numberOfEFoldingsStrongCouplingLKOct13}) is re-written as
\begin{equation}
	\label{eq:numberOfEFoldingsStrongCouplingLKOct13ReWrite}
	N = \frac{3\xi}{4}\int^{\phi_N}_{\phi_{\text{end}}}f'(\phi)\left(1+\frac{2f(\phi)}{3\xi[f'(\phi)]^2}-\frac{1}{\xi f(\phi)}\right)d\phi \,.
\end{equation}
The third term in the integrand is small compared to the first by the earlier condition $\xi f(\phi) \gg 1$, and it is assumed that the part of the potential being considered is not in the vicinity of any extrema which also prevents the second term being large \cite{LKOct13}. This second condition is equivalent to
\begin{equation}
	\label{eq:attractorCondition}
	3\left[\xi f'(\phi)\right]^2 > 2\xi f(\phi) \,.
\end{equation}

This expression is useful in determining the $\xi$ at which the attractor solution is approach and this idea is discussed in section 2.3. Following these steps $N$ is given by
\begin{equation}	
	N \simeq \frac{3}{4}\xi\left[f(\phi_N)-f(\phi_{\rm end})\right] \,,
\end{equation}
and $\epsilon$ by
\begin{equation}
	\label{eq:largeXiEpsilon}
	\epsilon \simeq \frac{4}{3\xi^2f^2(\phi)} \,.
\end{equation}
This expression for $\epsilon$ fixes $\xi f(\phi_{\rm end}) \sim 1 \ll N$ (incidentally showing that the strong-coupling assumption $\xi f(\phi) \gg 1$ will marginally fail towards the end of inflation) and so the field value at the end of inflation will contribute negligibly to the number of $e$-foldings giving the final expression for $N$ as
\begin{equation}
	\label{eq:numberOfEFoldingsStrongCouplingSimplifiedLKOct13}
	N \simeq \frac{3}{4}\xi f(\phi_N)  \,.
\end{equation}
This means that $f(\phi_N)$ is fixed in a potential independent manner and can be used in the expression for $\epsilon$ at large $\xi$.
A similar approach yields a potential-independent expression for $\eta$. This is most easily arrived at using
\begin{equation}
	\eta = 2\epsilon+\frac{\partial\epsilon}{\partial\varphi}\frac{1}{\sqrt{2\epsilon}} \,.
\end{equation}
In the strong-coupling limit the second term is
\begin{equation}
	\frac{\partial\epsilon}{\partial\varphi}\frac{1}{\sqrt{2\epsilon}} \simeq -\frac{4}{3\xi f(\phi)} \,.
\end{equation}
Taking the leading order of $1/N$ gives the Universal Attractor solution as \cite{LKOct13}:
\begin{equation}
	\label{eq:nSRHighCouplingLKOct13}
	r = \frac{12}{N^2} \quad ; \quad n_{\rm s} = 1-\frac{2}{N} \,.
\end{equation}
This is exactly the same point in the $n_{\rm s}$--$r$ plane as is given by the Starobinsky model (hereafter referred to as the `Starobinsky Point') and indicates a deep-seated connection between the two models which is fully explored in Ref.~\cite{Kehagias13}.

\subsection{The nature of the attractor solution}

In previous work on the Universal Attractor models the attractor point itself has been discussed in such a way as to either imply it coincides exactly with the Starobinsky point \cite{LKOct13} or state that it does \cite{Kehagias13}. However this is not true (e.g.\ in Figure 1 in Ref.~\cite{LKOct13} it can be seen that the numerically-generated trajectories converge to a point displaced from the Starobinsky point) and the Starobinsky point is actually only a first-order approximation to the true attractor point. The discrepancy stems from the simplification of the expression for the number of $e$-foldings.

The third term in the integrand of equation (\ref{eq:numberOfEFoldingsStrongCouplingLKOct13ReWrite}) is actually non-negligible. This can be seen by considering equation (\ref{eq:numberOfEFoldingsStrongCouplingSimplifiedLKOct13}) which fixes this term to be of order $1/N$ regardless of how large $\xi$ becomes. This means that, in the strong coupling regime, it is never completely negligible and when it is included the expression for $N$ becomes
\begin{equation}
	\label{eq:numberOfEFoldingsStrongCouplingMoreAccurateLKOct13}
	N = \frac{3}{4}\xi f(\phi_N)-\frac{3}{4}\text{log} \frac{f(\phi_N)}{f(\phi_{\text{end}})} \,.
\end{equation}
This extra term can be thought of as altering the precise number of e-foldings that are considered in the analysis, so that instead of $N$ the equations for the attractor point now contain $N+\delta N$. Following this modification through it is found that
\begin{equation}
	\label{eq:nSRPlaneCurve}
	\delta r = - \frac{24\delta N}{N^3} = -\frac{12}{N}\delta n_{\rm s} \,.
\end{equation}
So it is seen that the correction to $n_{\rm s}$ is of order $1/N^2$ and so $\epsilon$ must be included in the expression for $n_{\rm s}$ when considering the exact location of the Starobinsky point (to first order in slow-roll parameters). From equation (\ref{eq:nSRPlaneCurve}) it may be expected that the solutions would fall along this line with some dependence on the particular potential used. However $\delta N$ is in fact independent of $f(\phi)$. This can be seen by using an iterative approach. Taking $f(\phi_N)_1$ to be given by equation (\ref{eq:numberOfEFoldingsStrongCouplingSimplifiedLKOct13}), then
\begin{equation}
	f(\phi_N)_{i+1} = \frac{4}{3\xi}\left(N+\frac{3}{4}\text{log}\left[\frac{f(\phi_N)_i}{f(\phi_{\text{end}})}\right]\right) \,.
\end{equation}
This now does not depend on the particular form chosen for $f(\phi)$ since $f(\phi_{\rm end})$ is independent of the functional form of the potential.

\begin{figure}[t]
	\centering
		\includegraphics[width=0.8\textwidth]{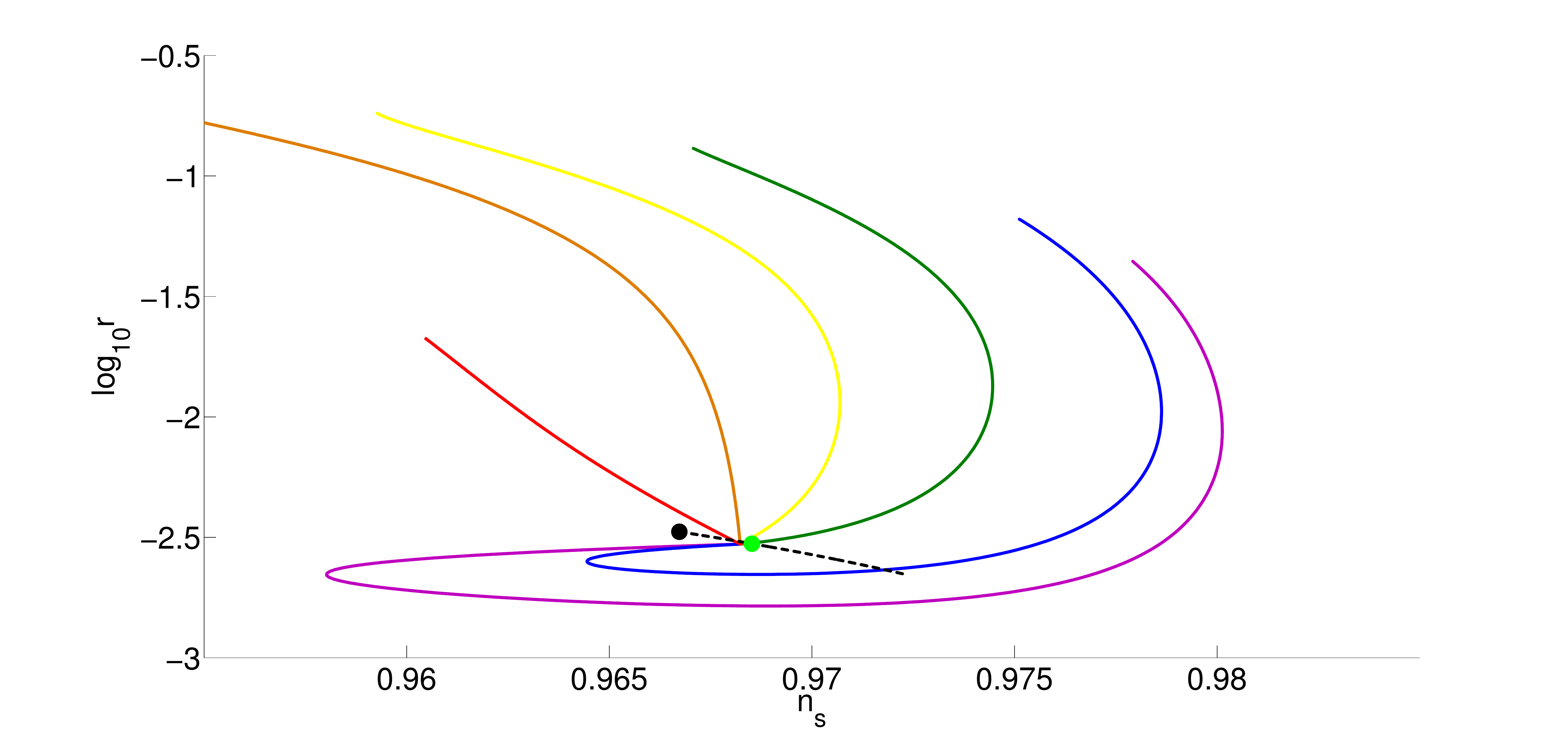}
		\caption{An illustration of the attractor behaviour for monomial potentials $V(\phi)=\phi^{\alpha}$ for \mbox{$\alpha = 6,4,3,2,1,2/3$} with $\alpha$ decreasing towards the blue end of the spectrum. The upper ends of the lines correspond to $\xi = 10^{-3}$. The black dot gives the point predicted by the Starobinsky model, the dashed line is given by equation (\ref{eq:nSRPlaneCurve}), and the green circle is the particular value on this line given by the potential-independent iterative approach.}
		\label{fig:attractorPlot}
\end{figure}

This shift of the attractor point can be seen in Figure \ref{fig:attractorPlot}. The trajectories are full numerical solutions to equation (\ref{eq:fullN}) and the green circle shows the point given by the potential-independent iterative method outlined above. The location of this circle becomes fixed very close to the attractor point even after a only a few iterations. This point is shifted away from the Starobinsky point, but given the accuracy of current data the shift is not significant. It could prove to be important if the uncertainty on both $n_{\rm s}$ and $N$ is reduced by a factor of approximately 10. If this regime became reality then the next order of slow-roll parameters may have to be considered in the calculation of $n_{\rm s}$ using \cite{stewart93}
\begin{equation}
	\label{eq:fullNs}
	\frac{n_{\rm s}-1}{2} = -3\epsilon+\eta-\frac{5+36C}{3}\epsilon^2+(8C-1)\epsilon\eta+\frac{1}{3}\eta^2-\frac{3C-1}{3}\xi^2_{\rm SR},
\end{equation}
where $C\simeq-0.73$ and $\xi_{\rm SR}$ is the third slow-roll parameter defined as
\begin{equation}
	\xi^2_{\rm SR} \equiv \frac{V'(\phi)V'''(\phi)}{V^2(\phi)}.
\end{equation}
From this the only $1/N^2$ contributions would come from the $\eta^2/3$ and the $\xi^2_{\rm SR}$ terms and these contributions are included in Figure \ref{fig:attractorPlot}. The effect of including these two terms is to shift the predicted $n_{\rm s}-r$ point for both the Starobinsky and Universal Attractor models but they do not contribute to the relative difference. There would also be next-order corrections to expression (\ref{eq:fullN}), due to the relation between the Hubble and potential slow-roll parameters. However we did not include these here as there is already a large uncertainty in the value which $N$ should take and these extra corrections are well within this range. If the accuracy to which $N$ is known was increased to the level suggested above, then the corrections become important and can be implemented using the expressions provided in Ref.~\cite{liddle:1994prd}.

\subsection{Approaching the attractor solution}

{\it Planck}'s non-detection of primordial tensor modes reignited interest in the Starobinsky model and those closely related to it, since these models suppress tensor modes and, as such, are placed at the `sweet spot' of the {\it Planck} data \cite{PlanckInflation}. The Universal Attractors become equivalent to the Starobinsky model in the large-coupling regime \cite{Kehagias13}, and so were generally considered where they were approaching the attractor behaviour.

\begin{figure}[t]
	\centering
		\includegraphics[width=0.45\textwidth]{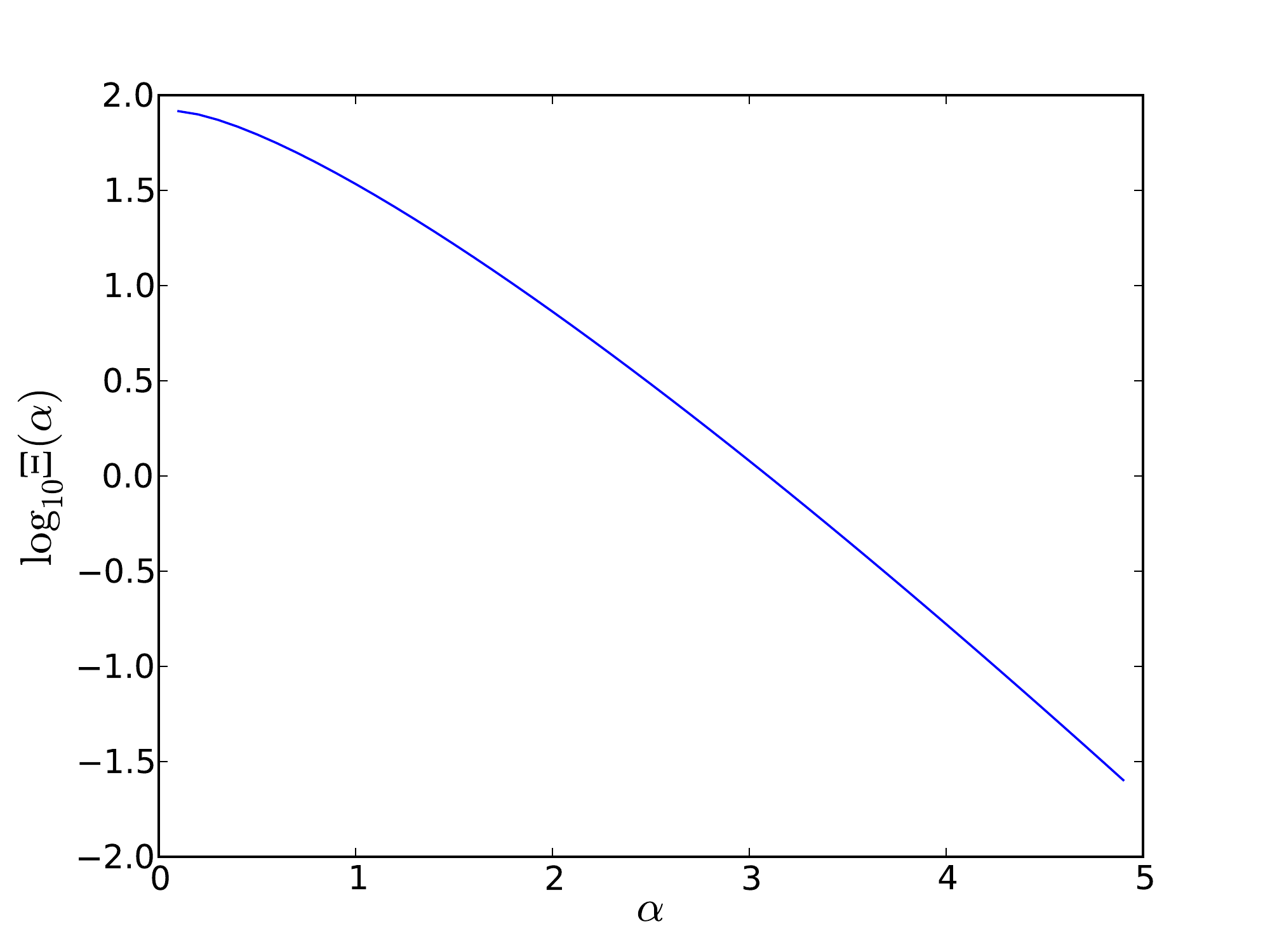}
		\includegraphics[width=0.45\textwidth]{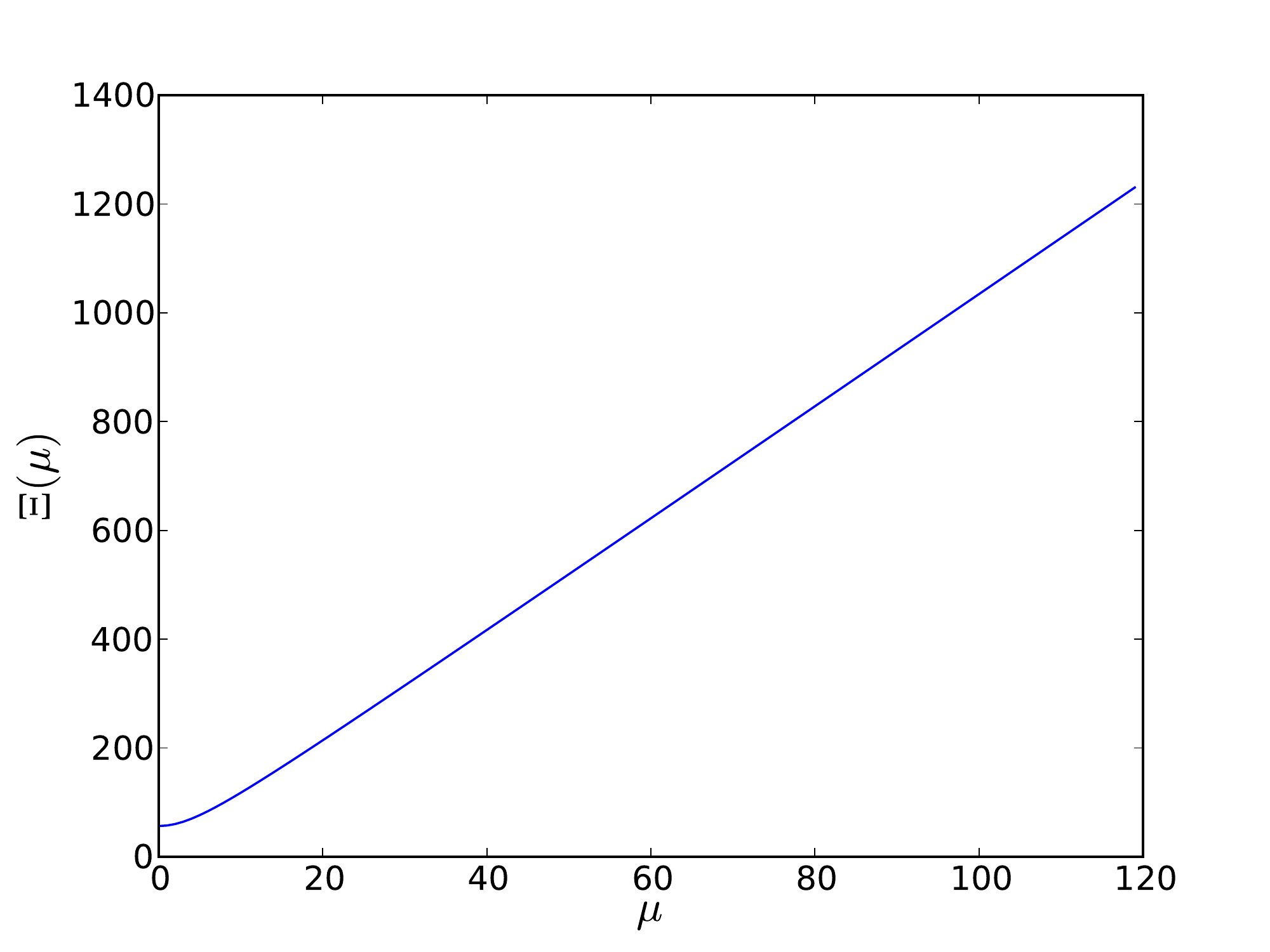}
\caption{The minimum coupling required to enter the attractor regime, encapsulated in the function $\Xi(\alpha)$ (for the monomial potential, left panel) and $\Xi(\mu)$ (for the natural inflation potential, right panel).}
		\label{fig:xiAlpha}
\end{figure}

The parameter values where this happens are encapsulated in equation (\ref{eq:attractorCondition}), which is required for the simplification of the expression for $\epsilon$ to equation (\ref{eq:largeXiEpsilon}) and is also the condition to neglect the second term in equation (\ref{eq:numberOfEFoldingsStrongCouplingLKOct13ReWrite}).
Thus, once a potential is specified the minimum $\xi$ required to be close to the attractor point is analytically calculable. In the case of the monomial potentials where $f(\phi) = \phi^{\alpha/2}$ then the minimum $\xi$ is a function of $\alpha$, $\Xi(\alpha)$, given by
\begin{equation}
	\label{eq:xiAlpha}
	\Xi(\alpha) \equiv \left(\frac{8}{3\alpha^2}\right)^{\alpha/4}\left(\frac{4N}{3}\right)^{1-\alpha/4} \,.
\end{equation} 
This function is plotted in Figure \ref{fig:xiAlpha} and can be seen to vary dramatically with $\alpha$.

A similar analysis can be carried out with Natural Inflation \cite{Adams:1992bn}, the other specific case considered in Ref.~\cite{LKOct13}. If the coupling function is taken to be $f(\phi) =  \sqrt{1+\cos\left(\phi/\mu\right)}$  then the minimum $\xi$ is now a function of $\mu$, $\Xi(\mu)$, given by 
\begin{equation}
	\Xi(\mu) \equiv \frac{4}{3}\sqrt{\frac{N^2}{2}+N\mu^2} \,.
\end{equation}
This function is also plotted in Figure \ref{fig:xiAlpha} and a comparison with the monomial case shows that the $\xi$ value required to approach the attractor is very different for different forms of the potential.

The general behaviour of $\Xi(\alpha)$ and $\Xi(\mu)$ is similar in that both show that for a more steep potential (higher $\alpha$ and lower $\mu$ respectively) the attractor is approached more rapidly, agreeing with the general statement of Ref.~\cite{LKOct13}.

In equation (\ref{eq:xiAlpha}) the expression for the number of e-foldings is taken to be the simple expression, equation (\ref{eq:numberOfEFoldingsStrongCouplingSimplifiedLKOct13}). Though this is only a first-order result, it suffices here to place an approximate lower bound on $\xi$.

\section{Observational constraints from BICEP2}

We now switch direction to consider the implications of the recent BICEP2 result \cite{BICEP2} for these models. Since at present the measured BICEP2 B-mode power spectrum has competing interpretations as due to primordial tensors or polarised dust, it is prudent to consider three scenarios, being that the signal is dominated by either one of those, or that the signal arises from a mixture of the two. If indeed BICEP2 is eventually attributed entirely to dust, we return to the situation after {\it Planck} where the Universal Attractor models sit favourably, and we have nothing to add to that case. We therefore consider in this section the possibilities that either all, or a significant fraction of all, of the observed B-mode signal has primordial tensor origin. In that case, observations have turned decisively against the Universal Attractor.

We see from the analysis at the end of the previous section that the Universal Attractor is attained typically only in the limit of high $\xi$, and in the past there has been no meaningful upper limit on this parameter. If BICEP2 is partly primordial, we now expect to find an upper limit on $\xi$ in different scenarios, the evaluation of which is the main objective of this section. The Universal Attractor models follow continuous trajectories in the $n_{\rm s}$--$r$ plane from the non-coupled case to the attractor point, meaning that they can cover a significant fraction of the new area of interest.

Before proceeding, we broaden the set of models under investigation. For the Universal Attractor model of equation (\ref{eq:jordanFrameLKOct13}), for completeness we now consider negative $\xi$ as well as positive $\xi$, in anticipation that $\xi = 0$ will not be excluded and hence $\xi$ can be bounded on either side. When $\xi$ is negative, the trajectories in the $n_{\rm s}$--$r$ plane move upwards from the minimally-coupled case, and hence are strongly constrained by {\it Planck}. Additionally, we consider the case where the non-minimal coupling is always of quadratic form, \mbox{$\Omega(\phi) = 1 + \xi \phi^2/2$}, rather than being related to the potential; such models have been widely discussed since the early days of inflation (for example Refs.~\cite{Salopek89,Kaiser:1994vs,Faraoni:1996rf}). In the case where the potential is quartic, then the quadratic coupling and Universal Attractor models coincide. For the quadratic-coupling model, a particular case of interest is the conformally-coupled case which is $\xi = -1/6$ in our conventions. In each case we focus on monomial chaotic inflation models, $V(\phi)=\phi^{\alpha}$. 

\begin{figure}[t]
	\centering
		\includegraphics[width=0.45\textwidth]{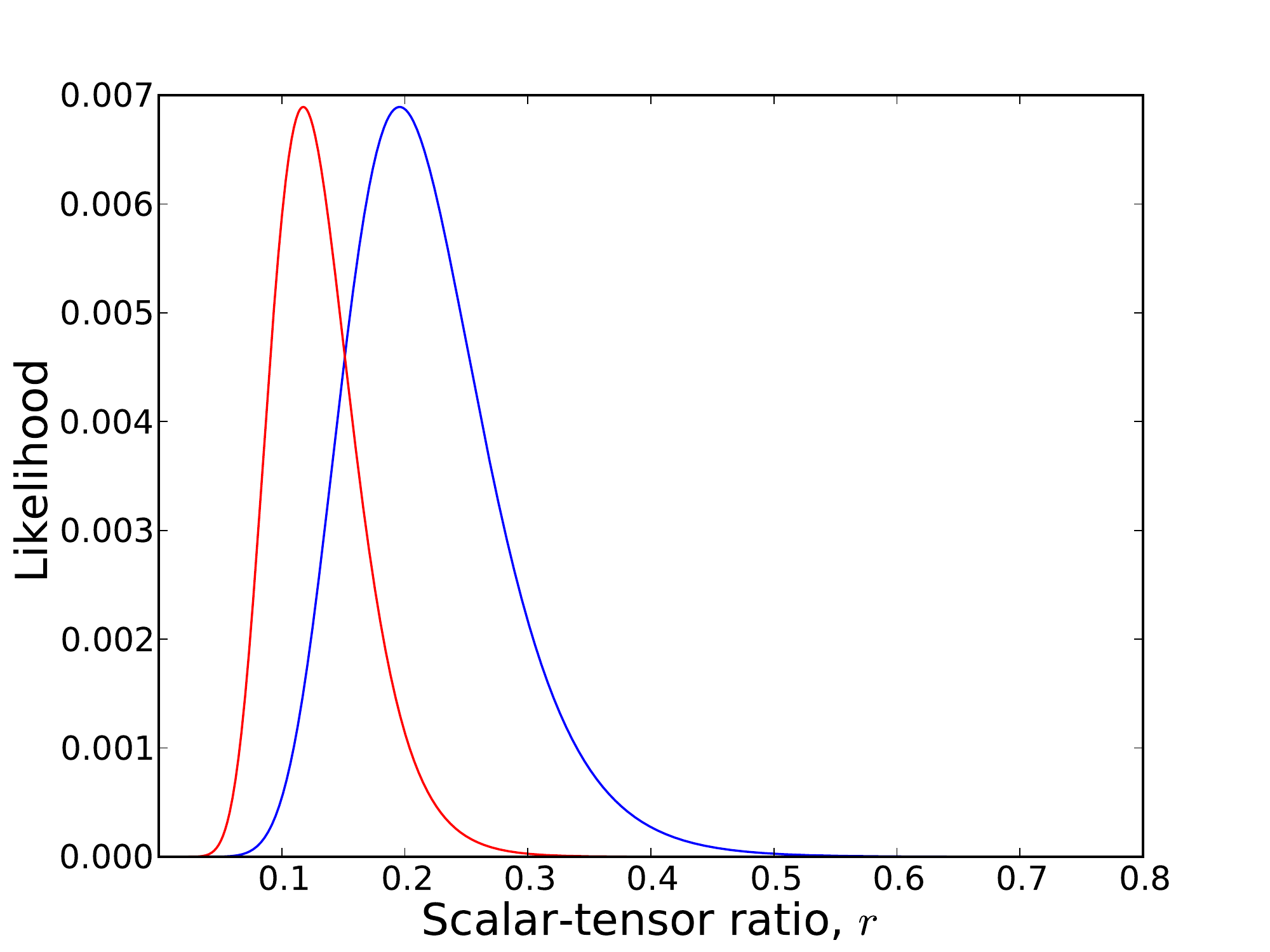}
		\includegraphics[width=0.45\textwidth]{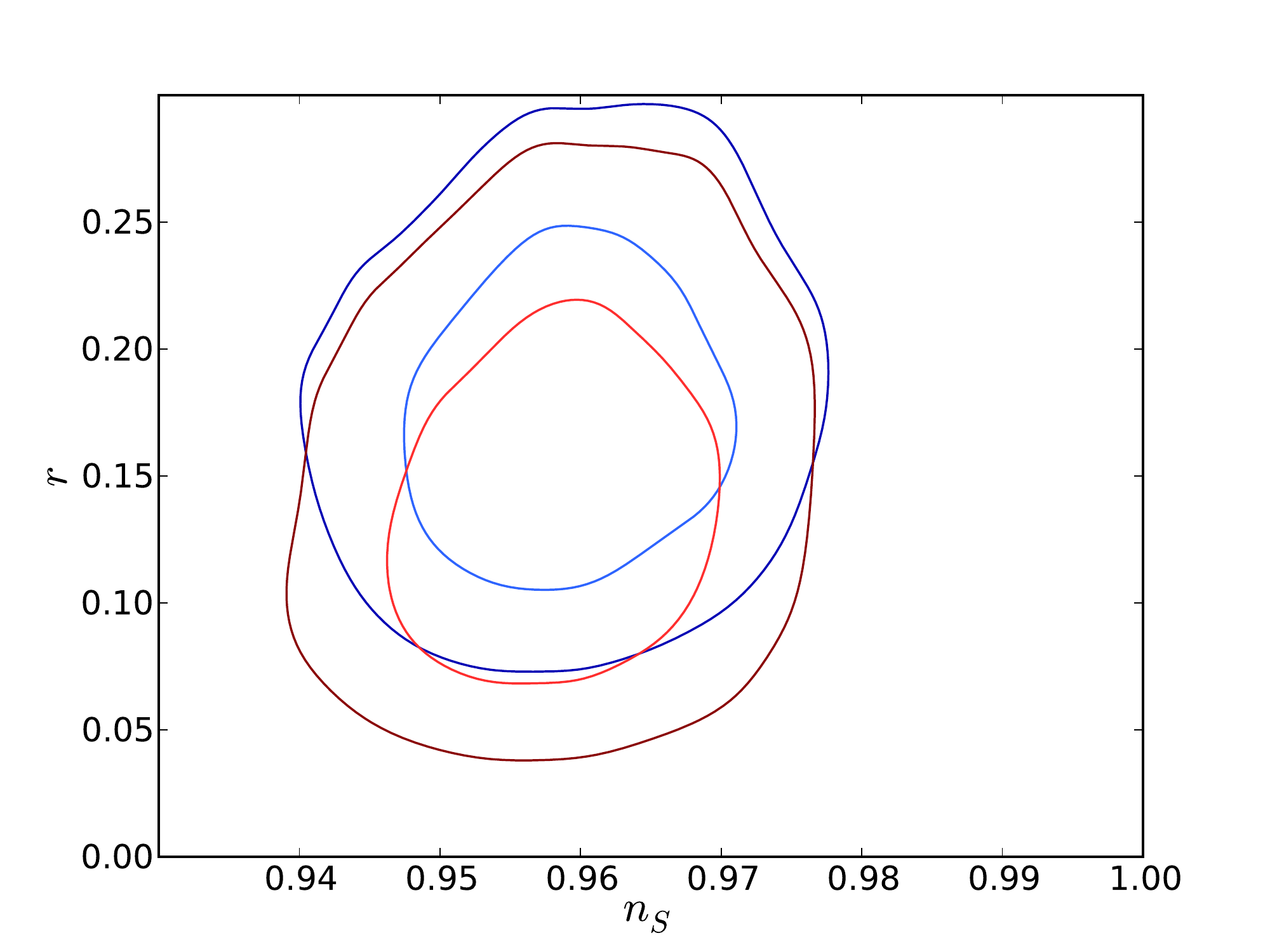}
		\caption{Left panel: the BICEP2 likelihood for $r$ (blue) and the modified likelihood where $r \rightarrow 0.6r$ (red). Right panel: the $68\%$ and $95\%$ confidence contours of the importance-sampled Planck MCMC chains using the unmodified BICEP2 likelihood (blue) and the modified likelihood (red).}
		\label{fig:bicepLikeMod}
\end{figure}

We use the BICEP2 data \cite{BICEP2} to importance sample the Markov Chain Monte Carlo (MCMC) chains provided by the {\it Planck} Collaboration \cite{PlanckInflation} so that an area in the $n_{\rm s}$--$r$ plane  corresponding to a $95\%$ confidence region is obtained, shown in blue in Figure \ref{fig:bicepLikeMod}. The results from BICEP2 are yet to be validated by other experiments and there are  unresolved questions over the extent of foreground effects which might be responsible for some or even all of the observed signal \cite{BICEP2,Lui14,Mortonson14,Flauger14}. As such it is important to consider the implications of alterations to the BICEP2 results on the above constraints. Specifically, Ref.~\cite{BICEP2} shows various foreground models which, to a good approximation, have the effect of rescaling the likelihood in $r$. Using this idea, importance sampling of the Planck Collaboration's MCMC chains for a modified likelihood, the left panel of Figure \ref{fig:bicepLikeMod}, corresponding to the strongest foreground model in Ref.~\cite{BICEP2}, gives another area in the $n_{\rm s}$--$r$ plane to consider, shown in red in the right panel of Figure \ref{fig:bicepLikeMod}. This area is actually not so different from the one obtained using the original BICEP2 likelihood, because the rescaled BICEP2 likelihood has less tension with the {\it Planck} one and so is less prone to be dragged down in the $r$ direction by it.

\begin{figure}[t]
	\centering
	\includegraphics[width=0.8\textwidth]{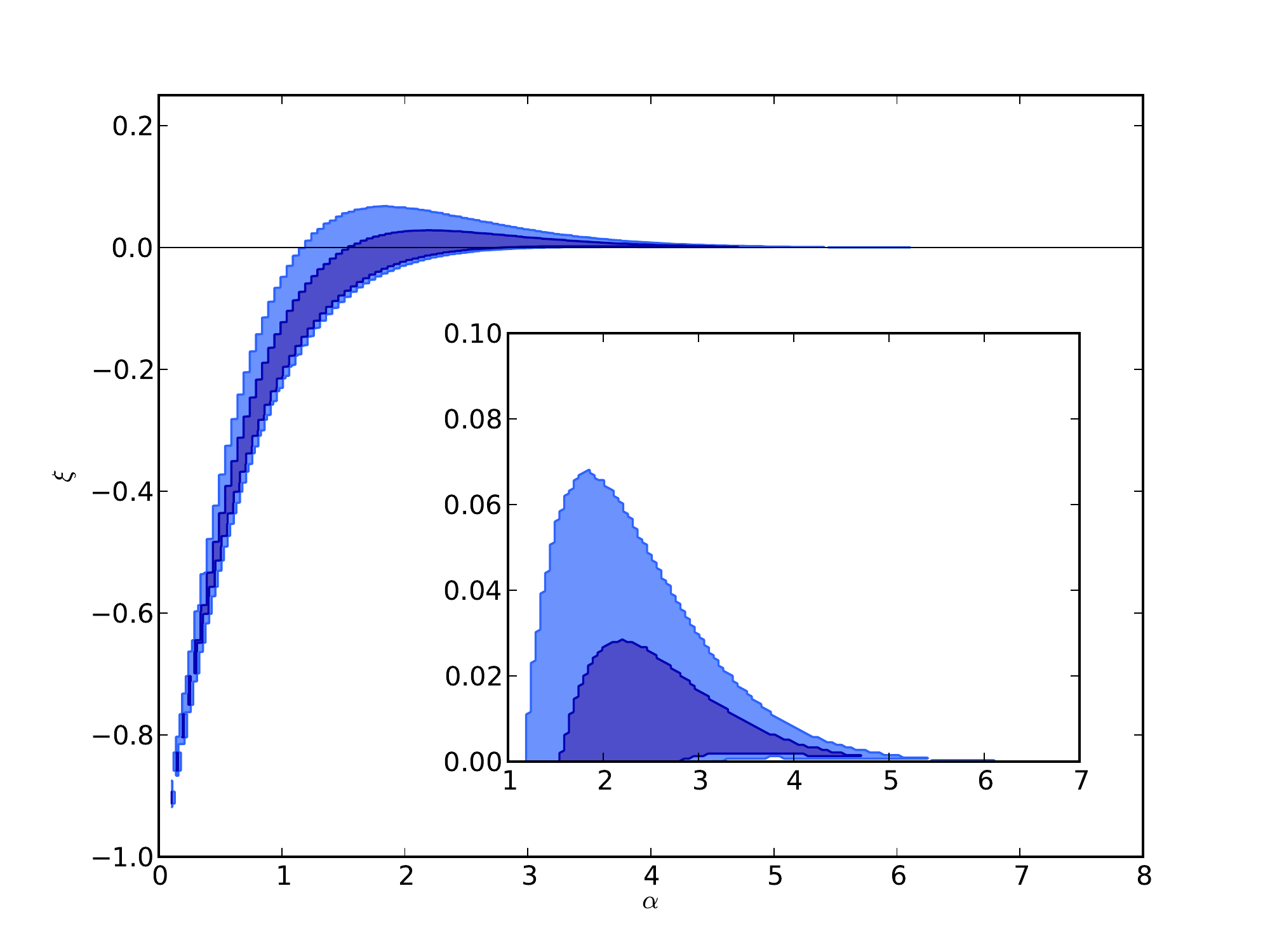}
	\caption{Constraints on the possible values of $\alpha$ and $\xi$ for a Universal Attractor model with a monomial potential using the raw BICEP2 likelihood with the dark blue indicating parameter combinations that give $n_{\rm s}$ and $r$ values inside the blue $68\%$ confidence region of Figure \ref{fig:bicepLikeMod}, and light blue those inside the $95\%$ confidence region.}
	\label{fig:allowedXiChaotic_100}
\end{figure}

The constraints placed on the Universal Attractors by the unaltered BICEP2 result are shown in Figure \ref{fig:allowedXiChaotic_100}. This gives an upper bound of $\xi < 0.07$ regardless of $\alpha$ and a lower bound, for well-motivated $\alpha$, of $\xi > -0.5$. For specific $\alpha$ values the bounds are even tighter, typically in a range of 0.1 or less. These are the first bounded constraints that have been possible for the Universal Attractor models.

\begin{figure}[t]
	\centering
	\includegraphics[width=0.8\textwidth]{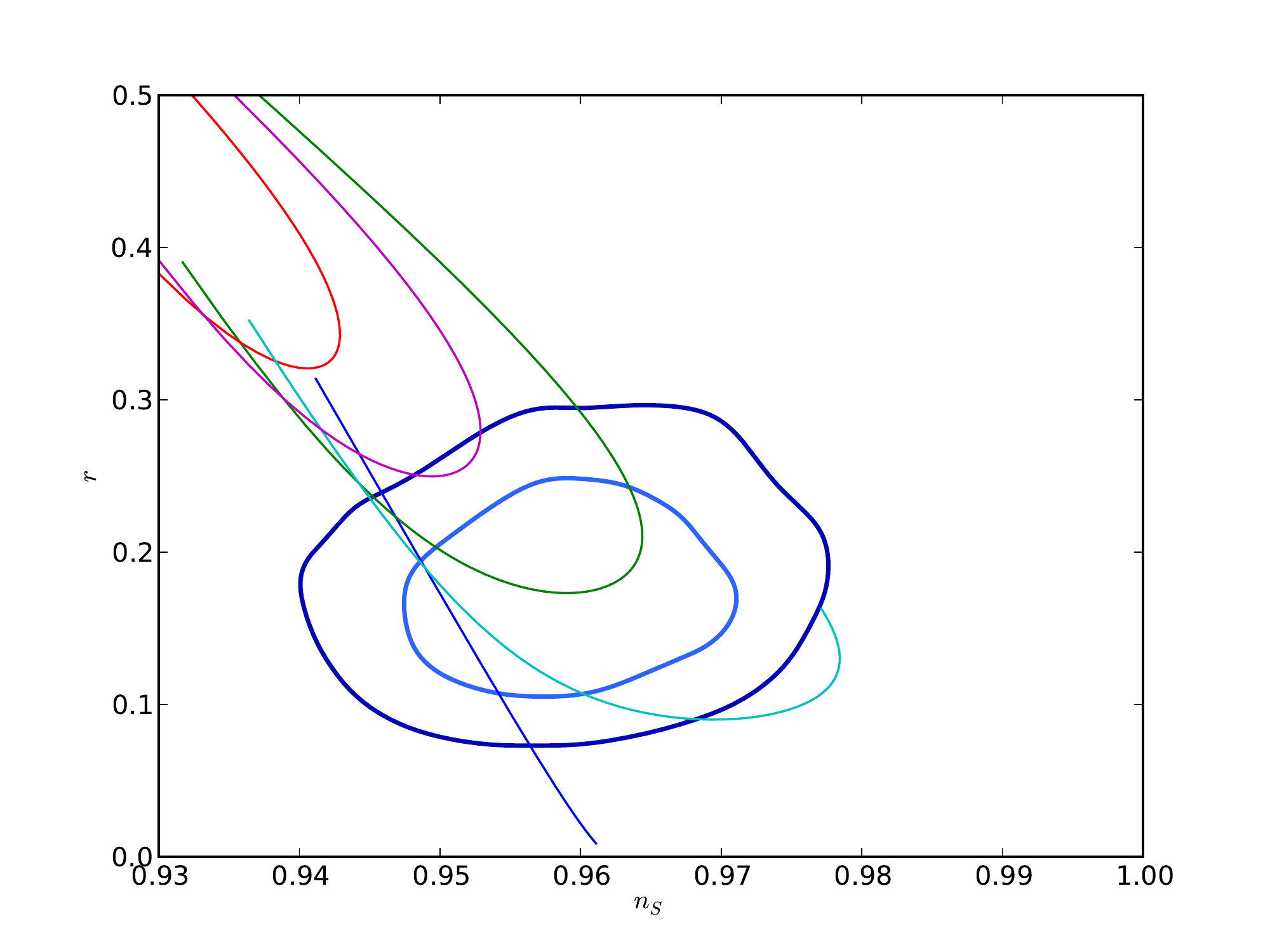}
	\caption{The trajectories followed in the $n_{\rm s}$--$r$ plane by the quadratically-coupled models, showing $\xi$ increasing from 0 to 0.15 for $\alpha = 4, 4.5, 5, 5.5, 6$ (represented by blue, cyan, green, purple and red respectively). There is a tail which loops round as $\xi$ is increased and can re-enter the allowed contours. The $\xi=0$ points are those given by the minimally coupled models and so are those at the end of the trajectories above and to the left of the observational contours.}
	\label{fig:confidenceContoursPlanckBicepTracts}
\end{figure}

\begin{figure}[t]
	\centering
	\includegraphics[width=0.8\textwidth]{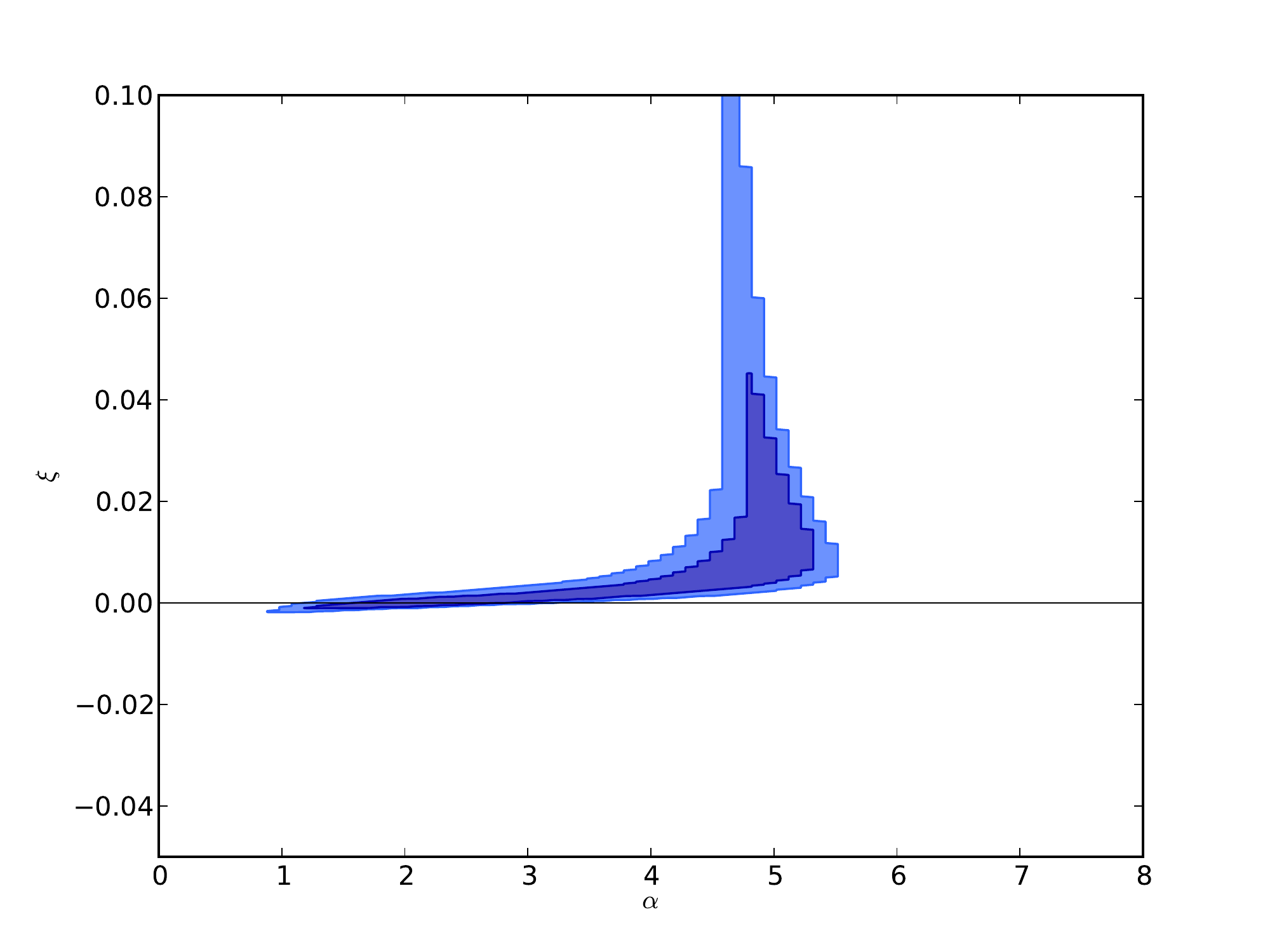}
	\caption{Constraints on the possible values of $\alpha$ and $\xi$ for an inflation model with quadratic non-minimal coupling and a monomial potential using the raw BICEP2 likelihood. The dark blue indicates parameter combinations that give $n_{\rm s}$ and $r$ values inside the blue $68\%$ confidence region of Figure \ref{fig:bicepLikeMod}, and light blue those inside the $95\%$ confidence region.}
	\label{fig:allowedXiFixed_100}
\end{figure}

Models with a quadratic coupling also follow trajectories in the $n_{\rm s}$--$r$ plane, Figure \ref{fig:confidenceContoursPlanckBicepTracts}, and so allow a similar analysis whose outcome is shown in Figure \ref{fig:allowedXiFixed_100}. The bounds for $\xi < 0$  at $\alpha < 4 $ are significantly tighter than in the Universal Attractor case, notably excluding the the conformal coupling case, $\xi=-1/6$, and are then broadly similar for $\alpha > 4$ so that the overall constraint on $\xi < 0$ is greatly improved. Just as in the case of the Universal Attractor it is not yet possible to completely constrain $\xi$ for arbitrary $\alpha$. Whereas for the Universal Attractor the tail that caused the problems was at very small $\alpha$ and so not problematic for well-motivated models, when considering the quadratic coupling the tail exists at $\alpha\rightarrow4^+$ and so cannot be ignored. This tail to infinity occurs because the trajectories begin to curl upwards at a certain value of $\xi$, seen in Figure \ref{fig:confidenceContoursPlanckBicepTracts}, and this value increases asymptotically as $\alpha\rightarrow4^+$. For larger values of $\alpha$ this flick-back occurs before the observational contours are ever reached, explaining why no models with $\alpha\gtrsim5.5 $ are allowed. Then decreasing $\alpha$ the flick-back now occurs inside the contours giving the wide range of allowed $\xi$ values at $\alpha\simeq4.5$. As $\alpha$ is decreased still further the flick-back occurs after the trajectory has passed through the observational contours meaning that the peak in Figure \ref{fig:allowedXiFixed_100} will fold back at higher $\xi$ values, going to infinity as $\alpha\rightarrow4^+$. In the case where $\alpha = 4$, the Universal Attractor model and the quadratic coupling model are exactly the same and these `flick-back' trajectories cease to exist and the constraints from both models are seen to be identical. Tighter observational constraints on $n_{\rm s}$ and $r$ may be able to resolve the issue of the tail to infinity if they rule out the flick-back part of the trajectories.

\begin{figure}[t]
	\centering
	\includegraphics[width=0.8\textwidth]{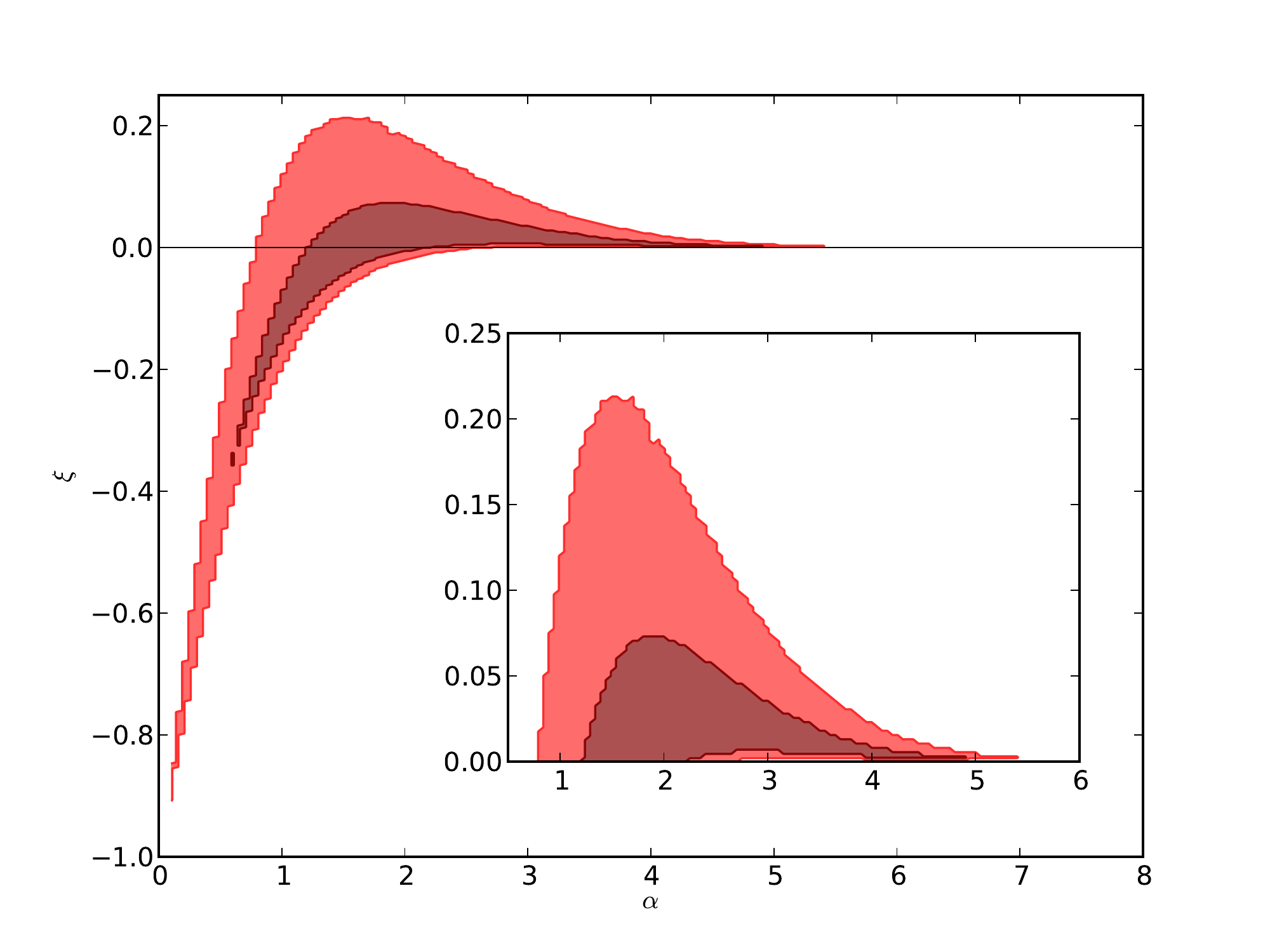}
	\caption{Constraints on the possible values of $\alpha$ and $\xi$ for a Universal Attractor with a monomial potential. Dark red indicates parameter combinations that give $n_{\rm s}$ and $r$ values inside the red $68\%$ confidence region of Figure \ref{fig:bicepLikeMod} and light red indicates those that are inside the $95\%$ confidence region.}
	\label{fig:allowedXiChaotic_60}
\end{figure}

\begin{figure}[t]
	\centering
	\includegraphics[width=0.8\textwidth]{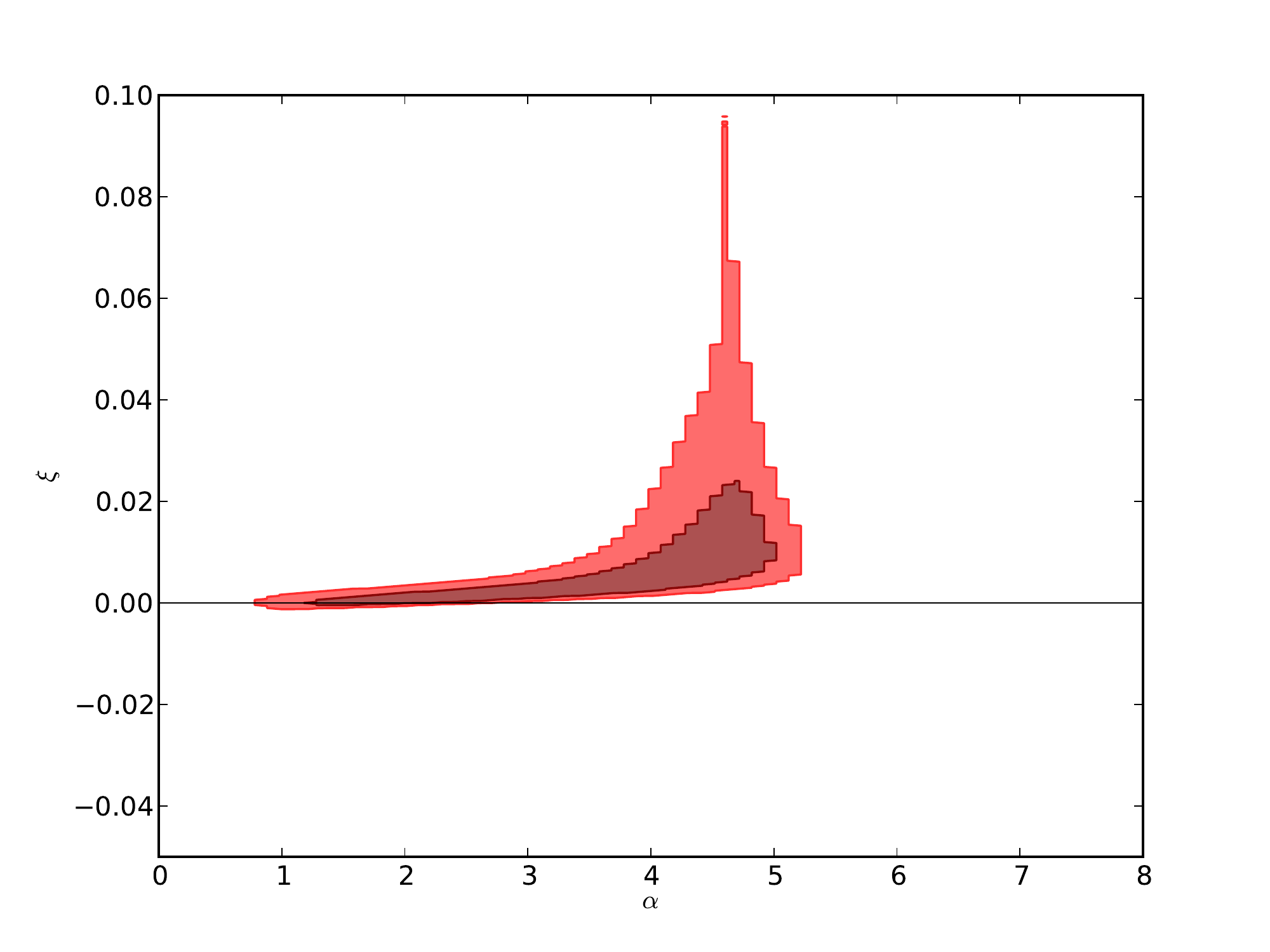}
	\caption{Constraints on the possible values of $\alpha$ and $\xi$ for an inflation model with quadratic non-minimal coupling and a monomial potential. Dark red indicates parameter combinations that give $n_{\rm s}$ and $r$ values inside the red $68\%$ confidence region of Figure \ref{fig:bicepLikeMod} and light red indicates those that are inside the $95\%$ confidence region.}
	\label{fig:allowedXiFixed_60}
\end{figure}

Even in the case of a significantly-reduced tensor contribution, as depicted in Figure~\ref{fig:bicepLikeMod}, the constraint on $\xi$ for the case of the Universal Attractors with a monomial potential is robust; Figure \ref{fig:allowedXiChaotic_60} shows an upper bound of $\xi < 0.22$ and a lower bound of $\xi > -0.5$. For the models with quadratic coupling the overall picture does not change substantially, with a slightly broader range of allowed $\xi$ for a given $\alpha$, Figure \ref{fig:allowedXiFixed_60}. The parameters are now constrained from above even in the limit $\alpha\rightarrow4^+$ since as $ \alpha=4 $ is approached the `flick-back' part of the tail begins to miss the shifted contours. This can be seen in Figure \ref{fig:allowedXiFixed_60} as there is now a truncated spike instead of the asymptotic behaviour seen from the unmodified BICEP2 analysis.

\section{Conclusions}

With the uncertainty over the interpretation of current cosmic microwave B-mode polarisation observations, we have decided to develop the Universal Attractor models in different directions to give a complete description of their current status. An in-depth analysis of the high-coupling limit of models shows a shift away from the Starobinsky point which has not been previously acknowledged. The exact positioning of the attractor point is currently a theoretical curiosity, but could be of interest if future CMB measurements improve sufficiently to constrain both $n_{\rm s}$ and $r$ by a further factor of ten.

The recent BICEP2 results suggest that attractor point is not favoured by observations and, as such, allows strong constraints to be placed on the Universal Attractors for the first time. For monomial potentials the magnitude of the coupling is restricted to $|\xi| < 1$ for well-motivated potentials by the BICEP2 result. This situation is not markedly changed when possible dust effects are taken into account, provided they are not too large.

Finally, the same analysis can be performed on models with a quadratic non-minimal coupling of the inflaton field. Again considering the Jordan frame potential to be a monomial the coupling is once again restricted to $|\xi| < 1$ for the majority of the possible powers. There is a tail to infinity as $\alpha \rightarrow 4^+$, which prevents a comprehensive bound being placed. However slight improvements in the observational constraints to $n_{\rm s}$ and $r$ could remove this entirely. Once again, allowing for a significant (though not dominant) dust contribution does not change the result, other than removing the tail to infinity due to a shift in the contours.

\acknowledgments
The authors would like to thank the referee for a thorough and useful report which helped improve the final paper. The authors were supported by STFC under grant numbers ST/K501980, ST/K006606/1 and ST/L000644/1.

\bibliographystyle{hieeetr}
\bibliography{universal}

\end{document}